\documentclass{pasj00}
 
\def\commenta{$^*$}
\def\commentb{$^\dagger$}
\def\commentc{$^\ddagger$}
\def\commentd{$^\S$}
\def\commente{$^\|$}
\def\commentf{$^\#$}

\def\inpress{in press}

\def\arxiv#1{ (arXiv astro-ph/#1)}

\DeclareAbbreviation\AAHam{Astron. Abh. Hamburg. Sternw.}
\DeclareAbbreviation\AARv{Astron. Astrophys. Rev.}
\DeclareAbbreviation\AAS{American Astron. Soc. Meeting Abstracts}
\DeclareAbbreviation\AcA{Acta Astron.}
\DeclareAbbreviation\actaa{Acta Astron.}
\DeclareAbbreviation\Afz{Astrofizika}
\DeclareAbbreviation\AGAb{Astronomische Gesellschaft Abstract Ser.}
\DeclareAbbreviation\an{Astron. Nachr.}
\DeclareAbbreviation\AnAp{Annales d'Astrophysique}
\DeclareAbbreviation\AnTok{Tokyo Astron. Obs. Annals, Sec. Ser.}
\DeclareAbbreviation\Ap{Astrophysics}
\DeclareAbbreviation\ARep{Astron. Rep.}
\DeclareAbbreviation\AstBu{Astrophys. Bull.}
\DeclareAbbreviation\ATel{Astron. Telegram}
\DeclareAbbreviation\ATsir{Astron. Tsirk.}
\DeclareAbbreviation\AcApS{Acta Astrophys. Sinica}
\DeclareAbbreviation\AstL{Astron. Lett.}
\DeclareAbbreviation\BaltA{Baltic Astron.}
\DeclareAbbreviation\BANS{Bull. of the Astron. Institutes of the Netherlands Suppl. Ser.}
\DeclareAbbreviation\BASI{Bull. Astron. Soc. India}
\DeclareAbbreviation\BeSN{Be Newslett.}
\DeclareAbbreviation\BHarO{Harvard Coll. Obs. Bull.}
\DeclareAbbreviation\CBET{Cent. Bur. Electron. Telegrams}
\DeclareAbbreviation\ChJAA{Chinese J. of Astron. and Astrophys.}
\DeclareAbbreviation\caa{Chinese J. of Astron. and Astrophys.}
\DeclareAbbreviation\CoAsi{Asiago Contr.}
\DeclareAbbreviation\CoSka{Contributions of the Astronomical Observatory Skalnat\'e Pleso}
\DeclareAbbreviation\GCN{GRB Coord. Netw. Circ.}
\DeclareAbbreviation\ErgAN{Erg. Astron. Nachr.}
\DeclareAbbreviation\ibvs{IBVS}
\DeclareAbbreviation\IEEEP{IEEE Proc.}
\DeclareAbbreviation\JAD{J. Astron. Data}
\DeclareAbbreviation\JApA{J. of Astrophys. and Astron.}
\DeclareAbbreviation\JAVSO{J. American Assoc. Variable Star Obs.}
\DeclareAbbreviation\JBAA{J. Br. Astron. Assoc.}
\DeclareAbbreviation\JPhCS{J. of Physics Conference Series}
\DeclareAbbreviation\JPSJ{J. Phys. Soc. Japan}
\DeclareAbbreviation\JSARA{J. of the Southeastern Assoc. for Research in Astron.}
\DeclareAbbreviation\LowOB{Lowell Obs. Bull.}
\DeclareAbbreviation\MitAG{Mitteil. der Astronom. Gesell. Hamburg}
\DeclareAbbreviation\MitVS{Mitteil. Ver\"{a}nderl. Sterne}
\DeclareAbbreviation\MmSAI{Mem. Soc. Astron. Ital.}
\DeclareAbbreviation\memsai{Mem. Soc. Astron. Ital.}
\DeclareAbbreviation\Msngr{Messenger}
\DeclareAbbreviation\NewA{New Astron.}
\DeclareAbbreviation\na{New Astron.}
\DeclareAbbreviation\NewAR{New Astron. Rev.}
\DeclareAbbreviation\nar{New Astron. Rev.}
\DeclareAbbreviation\NInfo{Nauchnye Informatsii}
\DeclareAbbreviation\NPhS{Nature Physical Science}
\DeclareAbbreviation\OAP{Odessa Astron. Publ.}
\DeclareAbbreviation\Obs{Observatory}
\DeclareAbbreviation\OEJV{Open Eur. J. on Variable Stars}
\DeclareAbbreviation\PASA{Publ. Astron. Soc. Australia}
\DeclareAbbreviation\PASAu{Publ. Astron. Soc. Australia}
\DeclareAbbreviation\PAZh{Pis'ma AZh}
\DeclareAbbreviation\PJAB{Proc. Japan Acad. Ser. B}
\DeclareAbbreviation\POBeo{Publ. de l'Observatoire Astronomique de Beograd}
\DeclareAbbreviation\PCCP{Phys. Chem. Chem. Phys.}
\DeclareAbbreviation\PhR{Phys. Rep.}
\DeclareAbbreviation\PVSS{Publ. Variable Stars Sect. R. Astron. Soc. New Zealand}
\DeclareAbbreviation\PZ{Perem. Zvezdy}
\DeclareAbbreviation\PZP{Perem. Zvezdy, Prilozh.}
\DeclareAbbreviation\QJRAS{QJRAS}
\DeclareAbbreviation\RA{Ricerche Astronomiche}
\DeclareAbbreviation\RMxAA{Rev. Mexicana Astron. Astrof.}
\DeclareAbbreviation\RvMA{Reviews of Modern Astron.}
\DeclareAbbreviation\SASS{Society for Astronom. Sciences Ann. Symp.}
\DeclareAbbreviation\Sci{Science}
\DeclareAbbreviation\SPIE{SPIE Proc.}
\DeclareAbbreviation\SvA{Soviet Astronomy}
\DeclareAbbreviation\SvAL{Soviet Astronomy Letters}
\DeclareAbbreviation\VeSon{Ver\"{o}ff. Sternw. Sonneberg}
\DeclareAbbreviation\VSOLJBul{VSOLJ Variable Star Bull.}
\DeclareAbbreviation\yCat{VizieR Online Data Catalog}
\DeclareAbbreviation\ZA{Z. Astrophys.}

\def\PublisherCambridge{Cambridge: Cambridge University Press}

\def\PublisherSpringer{Berlin: Springer-Verlag}

\begin{document}
\SetRunningHead{T. Kato and Y. Osaki}{Estimation of Binary Mass Ratios Using Superhumps}

\Received{201X/XX/XX}
\Accepted{201X/XX/XX}

\title{New Method to Estimate Binary Mass Ratios by Using Superhumps}

\author{Taichi \textsc{Kato}}
\affil{Department of Astronomy, Kyoto University,
       Sakyo-ku, Kyoto 606-8502}
\email{tkato@kusastro.kyoto-u.ac.jp}

\and

\author{Yoji \textsc{Osaki}}
\affil{Department of Astronomy, School of Science, University of Tokyo,
Hongo, Tokyo 113-0033}
\email{osaki@ruby.ocn.ne.jp}


\KeyWords{accretion, accretion disks
          --- stars: dwarf novae
          --- stars: novae, cataclysmic variables
         }

\maketitle

\begin{abstract}
We propose a new dynamical method to estimate binary mass ratios by 
using the period of superhumps in SU UMa-type dwarf novae during
the growing stage (the stage A superhumps). This method is based on a 
working hypothesis in which the period of the superhumps at 
the growing stage is determined by the dynamical precession rate 
at the 3:1 resonance radius, a picture suggested in our 
new interpretation of the superhump period 
evolution during the superoutburst \citep{osa13v344lyrv1504cyg}. 
By comparison with the objects
with known mass ratios, we show that our method can provide
sufficiently accurate mass ratios comparable to those obtained
by quiescent eclipse observations.  This method is very advantageous
in that it requires neither eclipses, nor an experimental
calibration.  It is particularly suited for exploring 
the low mass-ratio end of the evolution of cataclysmic variables,
where the secondary is undetectable by conventional methods.
Our analysis suggests that previous estimates of mass ratios
using superhump periods during superoutburst were systematically
underestimated for low mass-ratio systems and we provided a new
calibration.  It suggests that most of WZ Sge-type
dwarf novae have secondaries close to the border of the lower
main-sequence and brown dwarfs, and most of the objects have not
yet reached the evolutionary stage of period bouncers.
Our result is not in contradiction with an assumption that 
the observed minimum period ($\sim$77 min) of ordinary hydrogen-rich
cataclysmic variables is indeed the period minimum.
We highlight the importance of early observation of stage A 
superhumps and propose a future desirable strategy of observation.
\end{abstract}

\section{Introduction}

   Cataclysmic variables (CVs) are close binary systems composed of
a white dwarf primary and a red-dwarf or brown-dwarf secondary
transferring matter via Roche overflow [for a review of CVs,
see \citet{war95book}; \citet{hel01book}].  The transferred matter
forms an accretion disk around the white dwarf.  Dwarf novae
(DNe) are a class of CVs that undergo recurrent outbursts,
whose origin is believed to be the thermal instability in
the accretion disk.  SU UMa-type dwarf novae are a subclass of
DNe that show long-lasting outbursts called superoutbursts and
superhumps during superoutbursts.  The origin of superhumps
is generally believed to be the tidal instability \citep{whi88tidal}
generated at the radius in 3:1 resonance with the orbital motion of
the secondary.  In the thermal-tidal instability (TTI) model
\citep{osa89suuma}, the superoutburst is triggered by an outburst
which produces an expansion of the accretion disk to the radius
of the 3:1 resonance and thereby excites the tidal instability
to produces superhumps\footnote{
  \citet{sma09SH} has proposed an alternative model to the
  standard model of the tidal dissipation of the eccentric precessing disk
  for the superhump light source based on the enhanced mass
  transfer (EMT) model, in which the superhump may be produced
  by variable hot-spot brightness due to variable mass transfer
  rate which is in turn produced by periodic variation in irradiation heating
  of the secondary star. Our discussion presented in this paper
  can not be applied to Smak's EMT model. This is firstly because the basic
  underlying ``clock'' for variable irradiation in his model is rather vague
  as \citet{sma09SH} has just mentioned it as ``a tidal origin''
  and secondly because his model involves the time delay between enhanced
  irradiation heating on the surface of the secondary and enhanced mass
  transfer, $\Delta t_{\rm flow}$, and this time delay is very uncertain
  because of uncertain nature of the hydrodynamic flow
  on the surface of the secondary.
} and strong tidal removal of the angular
momentum from the disk [for a review, see \citet{osa96review}].

   There is a long history to try to extract the elusive 
mass ratio $q=M_2/M_1$ 
of the binary from the fractional superhump (SH) excess in period,  
which is defined by $\varepsilon \equiv P_{\rm SH}/P_{\rm orb}-1$ 
where $P_{\rm SH}$ and $P_{\rm orb}$ are the superhump period 
and the orbital period of the binary, respectively. That is to 
try to estimate the binary's mass ratio $q$ from easily observable 
superhump-period excess $\varepsilon$.  It is convenient 
to introduce another expression, $\varepsilon^*$, the apsidal precession rate 
of the eccentric disk, $\omega_{\rm pr}$, over the binary orbital angular 
frequency, $\omega_{\rm orb}$, which is written as  
$\varepsilon^* \equiv \omega_{\rm pr}/\omega_{\rm orb}=1-P_{\rm orb}/P_{\rm SH}$. 
These two $\varepsilon$'s are related with each other by 
$\varepsilon^*=\varepsilon/(1+\varepsilon)$ or $\varepsilon=\varepsilon^*/(1-\varepsilon^*)$. 

   The apsidal precession rate of an eccentric disk in SU UMa stars was 
first discussed by \citet{osa85SHexcess}, who showed that 
the precession rate is a function 
of the binary's mass ratio, $q$, and the disk radius $R_d$ because 
the apsidal precession rate of an accretion disk is basically determined 
by the gravitational tidal torques of the secondary star acting 
on the eccentric disk. 
Thus it was natural to use $\varepsilon^*$ or $\varepsilon$ 
for estimating $q$. This method was thus used to estimate the mass ratio 
for X-ray binaries (\cite{min92BHXNSH}), and for cataclysmic variable 
stars (\cite{pat98evolution}; \cite{pat05SH} and refinements 
by various authors).

   However, there are two problems in this approach. One is that 
the superhump period varies with time during a superoutburst and thus 
a question naturally arise at what stage we should apply 
the $\varepsilon-q$ relation.  The other problem is that the 
the pressure effects within the disk contribute 
in determining the apsidal precession rate of the eccentric disk 
besides the pure dynamical effects of the secondary tidal torques 
and thus it is not a pure dynamical problem. 

\begin{figure}
  \begin{center}
    \FigureFile(88mm,110mm){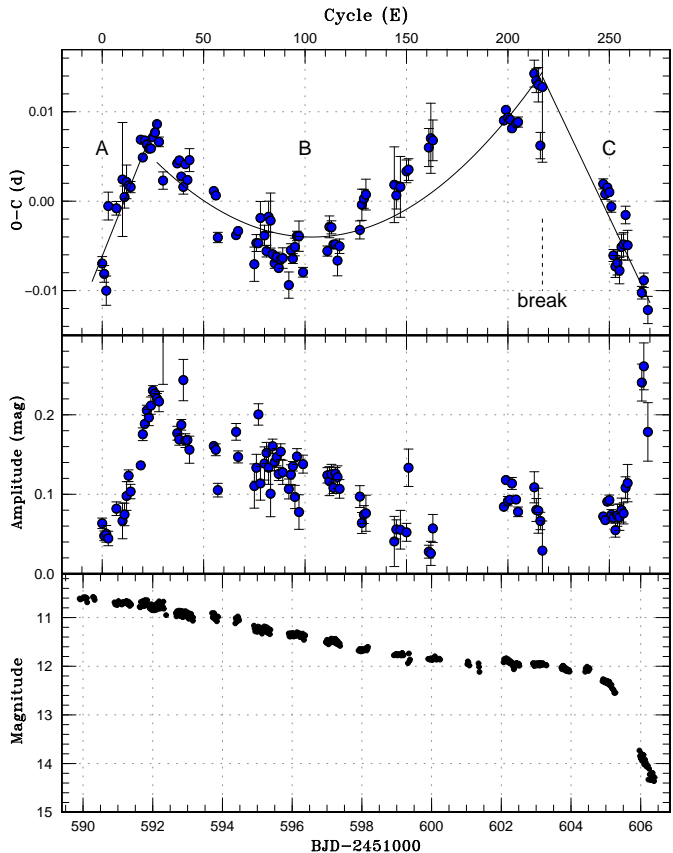}
  \end{center}
  \caption{Representative $O-C$ diagram showing three stages (A--C)
  of $O-C$ variation.  The data were taken from the 2000 superoutburst
  of SW UMa.  (Upper:) $O-C$ diagram.  Three distinct stages
  (A -- evolutionary stage with a longer superhump period, 
  B -- middle stage, and C -- stage after
  transition to a shorter period) and the location of the period break
  between stages B and C are shown. (Middle): Amplitude of superhumps.
  During stage A, the amplitude of the superhumps grew.
  (Lower:) Light curve.}
  \label{fig:stagerev}
\end{figure}

   As for the first problem, through an extensive survey of SU UMa-type 
dwarf novae, \citet{Pdot} have shown that the superhump periods in many
SU UMa-type dwarf novae exhibit a characteristic pattern of variation where 
their $O-C$ diagrams show a variation with the stage A--B--C:
(1) stage A with growing superhumps; the superhump period is longer
than in other stages, (2) stage B with a shorter superhump period;
in objects with short $P_{\rm orb}$, the period  derivative
$P_{\rm dot} \equiv \dot{P}/P$ is positive, i.e. the superhump
period increases during this stage, (3) stage C with a shorter
superhump period than in stage B; during stage C, the period
is relatively constant and this stage often continues after
the termination of the superoutburst (figure \ref{fig:stagerev}).

   \citet{Pdot} experimentally assumed that the period
of the superhumps (or the fractional superhump excess) at the
beginning of stage B reflects the precession rate at the 3:1
resonance.  This selection was based on the finding that
this period is very close to the superhump stable period of stage C
in objects with $P_{\rm dot} > 0$.  \citet{Pdot} regarded
stage A superhumps as immature superhumps, and the fully mature
superhump at the beginning of stage B reflects the fully grown
3:1 resonance.  Using this assumption and using the dependence
of the dynamical precession rate on the radius, \citet{Pdot}
estimated the disk radius variation in a purely dynamical way.
This treatment, however, neglected the pressure effect.

   As for the second problem of the pressure effects, it is 
known that they act to reduce the apsidal precession rate of 
the eccentric disk (\cite{lub92SH}; \cite{hir93SHperiod};
\cite{mur98SH}; \cite{mon01SH}; \cite{pea06SH}).
Among them \citet{pea06SH} showed that the pressure effect
is necessary to express the observation, and derived the
strength of the pressure effect for the calibration objects
in \citet{pat05SH} using the formulation by \citet{lub92SH}.

   In this paper, we propose a new dynamical method to 
estimate binary mass ratios by using the period of superhumps in 
SU UMa-type dwarf novae during the growing stage 
(the stage A superhumps), and examine the periods of superhumps 
recorded in \citet{Pdot}, \citet{Pdot2}, \citet{Pdot3}, \citet{Pdot4}.
In section \ref{sec:model}, we present the major premise of our working 
model and its formulation. 
In section \ref{sec:stagea} we present  a test for the interpretation 
using a comparison with systems with known $q$ and a comparison of the resultant
evolutionary sequence. In section \ref{sec:discussion}
we present various applications and implications.

\section{The Main Premise of Our Working Model and Its Formulation}
\label{sec:model} 

   \citet{osa13v344lyrv1504cyg} have proposed a new interpretation 
for the time evolution of the superhump period during superoutbursts of 
V344 Lyr and V1504 Cyg in the Kepler data, in particular by using 
a comparison of simultaneously recorded positive and negative superhumps. 
In SU UMa stars, the longest superhump period (and therefore the highest 
apsidal precession rate) occurs at the growing stage of the superhump 
(stage A in \cite{Pdot} classification).  \citet{osa13v344lyrv1504cyg} 
have interpreted that this is most likely given by that 
corresponding to the dynamical precession rate at the 3:1 resonance radius.
Its following rapid decrease in the SH period (a transition from 
stage A to stage B) is then understood as due to propagation of 
the eccentricity wave to the inner part of the disk by which 
a larger portion of the disk is involved in determining 
the precession rate of the whole disk. 

   In this interpretation, the dynamical precession rate at
the 3:1 resonance is represented in the growing stage 
of superhumps (stage A) when the eccentric wave is still confined to 
the location of the resonance.  In this paper we adopt this 
interpretation as our working model and we examine its consequence below.   
This interpretation leads to an important consequence: 
$q$ can be directly determined (without an experimental
coefficient) by a dynamical way from $\varepsilon^*$ of
the stage A superhumps.  The traditional way of using
$\varepsilon^*$ for stage B superhumps to estimate $q$
(e.g. \cite{pat05SH}; \cite{Pdot}) suffers from the unknown
pressure effect, which is expected to be the strongest
in stage B, resulting large uncertainties.

   The dynamical precession rate, $\omega_{\rm dyn}$, at radius $r$ 
in the disk can be expressed by (see, \cite{hir90SHexcess})
\begin{eqnarray}
\label{equ:presfreq}
\omega_{\rm dyn}/\omega_{\rm orb} & = \frac{q}{\sqrt{1+q}}\Bigl[\frac{1}{4}\frac{1}{\sqrt{r}}\frac{d}{dr}\Bigr(r^2\frac{db_{1/2}^{(0)}}{dr}\Bigr)\Bigr] \nonumber \\
& = \frac{q}{\sqrt{1+q}} \Bigl[\frac{1}{4}\frac{1}{\sqrt{r}} b_{3/2}^{(1)}\Bigr].
\end{eqnarray}
where $r$ is the dimensionless radius measured in units of the binary 
separation $A$, $\omega_{\rm orb}$ is the angular frequency of 
the binary motion,  
and $\frac{1}{2}b_{s/2}^{(j)}$ is the Laplace coefficient
\begin{equation}
\label{equ:laplace}
\frac{1}{2}b_{s/2}^{(j)}(r)=\frac{1}{2\pi}\int_0^{2\pi}\frac{\cos(j\phi)d\phi}
{(1+r^2-2r\cos\phi)^{s/2}},
\end{equation}

   We estimate the dynamical precession rate at the 3:1 resonance radius, 
which is given by 
\begin{equation}
\label{equ:radius31}
r_{3:1}=3^{(-2/3)}(1+q)^{-1/3}.
\end{equation} 

\begin{figure}
  \begin{center}
    \FigureFile(88mm,70mm){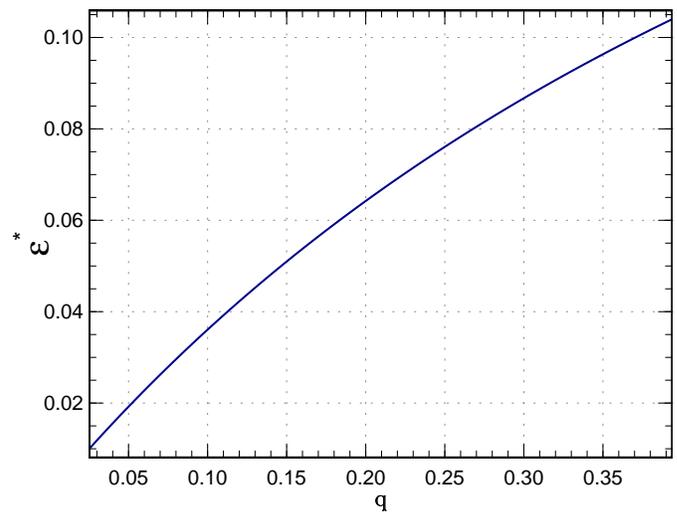}
  \end{center}
  \caption{Relation between $\varepsilon^*$ of stage A and $q$.}
  \label{fig:qeps31}
\end{figure}

   We can now calculate the dynamical precession rate of the eccentric 
disk at the 3:1 resonance radius and we show our results 
in table \ref{tab:stageaepsq} and figure \ref{fig:qeps31}
for $\varepsilon^*-q$ relation.  We also give approximate
analytic formulae as follows:
\begin{equation}
\label{equ:qtoepspoly}
\varepsilon^* = 0.00027 + 0.402q - 0.467q^2 + 0.297q^3,
\end{equation} 
which has a maxmimum error 0.00004 in $\varepsilon^*$ and
\begin{equation}
\label{equ:epstoqpoly}
q = -0.0016 + 2.60\varepsilon^* +3.33(\varepsilon^*)^2 + 79.0(\varepsilon^*)^3,
\end{equation} 
which has a maxmimum error 0.0004 in $q$, respectively,
in the range of $0.025 \le q \le 0.394$.
Since numerical integration using equation (\ref{equ:laplace})
converges sufficiently quickly, we recommend to use integration
rather than polynomial approximations.

   By identifying the observed $\varepsilon^*$ for stage A superhumps 
to the dynamical precession rate at the 3:1 resonance radius, 
we can obtain $q$.  This is our basic strategy in this paper.  

\begin{table}
\caption{Relation between $\varepsilon^*$ of stage A and $q$.}
\label{tab:stageaepsq}
\begin{center}
\begin{tabular}{cc|cc|cc}
\hline
$\varepsilon^*$ & $q$ & $\varepsilon^*$ & $q$ & $\varepsilon^*$ & $q$ \\ 
\hline
0.010 & 0.025 & 0.042 & 0.119 & 0.074 & 0.241 \\
0.012 & 0.030 & 0.044 & 0.126 & 0.076 & 0.250 \\
0.014 & 0.036 & 0.046 & 0.133 & 0.078 & 0.259 \\
0.016 & 0.041 & 0.048 & 0.140 & 0.080 & 0.268 \\
0.018 & 0.047 & 0.050 & 0.147 & 0.082 & 0.277 \\
0.020 & 0.052 & 0.052 & 0.154 & 0.084 & 0.287 \\
0.022 & 0.058 & 0.054 & 0.161 & 0.086 & 0.296 \\
0.024 & 0.064 & 0.056 & 0.168 & 0.088 & 0.306 \\
0.026 & 0.069 & 0.058 & 0.176 & 0.090 & 0.317 \\
0.028 & 0.075 & 0.060 & 0.183 & 0.092 & 0.327 \\
0.030 & 0.081 & 0.062 & 0.191 & 0.094 & 0.337 \\
0.032 & 0.087 & 0.064 & 0.199 & 0.096 & 0.348 \\
0.034 & 0.093 & 0.066 & 0.207 & 0.098 & 0.359 \\
0.036 & 0.100 & 0.068 & 0.215 & 0.100 & 0.370 \\
0.038 & 0.106 & 0.070 & 0.224 & 0.102 & 0.382 \\
0.040 & 0.113 & 0.072 & 0.232 & 0.104 & 0.394 \\
\hline
\end{tabular}
\end{center}
\end{table}

   We now compare our interpretation with the smoothed particle 
hydrodynamics (SPH) simulations of superhumps by \citet{mur98SH}, 
who demonstrated the evolution of the superhump period for one of his 
simulations in a form of $O-C$ diagram.  His $O-C$ diagram 
showed a pattern similar to stage A--B transition during
the growing stage of superhumps. 
Our result is now compared with SPH result
[\citet{mur98SH}, $P_{\rm SH}/P_{\rm orb}$ (max) corresponding to stage A].
The agreement is fair: $\varepsilon^*$=0.090 for $q$=0.25,
$\varepsilon^*$=0.065--0.069 for $q$=3/17=0.176 and
$\varepsilon^*$=0.049 for $q$=1/9=0.111.
Considering the intrinsic difficulty in measuring the periods
from SPH simulations, the agreement appears to be sufficient. 

   Since the SPH simulations are favorably compared with our 
interpretation, let us turn to observations. 

\section{Comparison with Observations}\label{sec:stagea}

\subsection{Stage A Superhumps in Systems with Known Mass Ratios}
\label{sec:knownq}

\begin{figure}
  \begin{center}
    \FigureFile(88mm,88mm){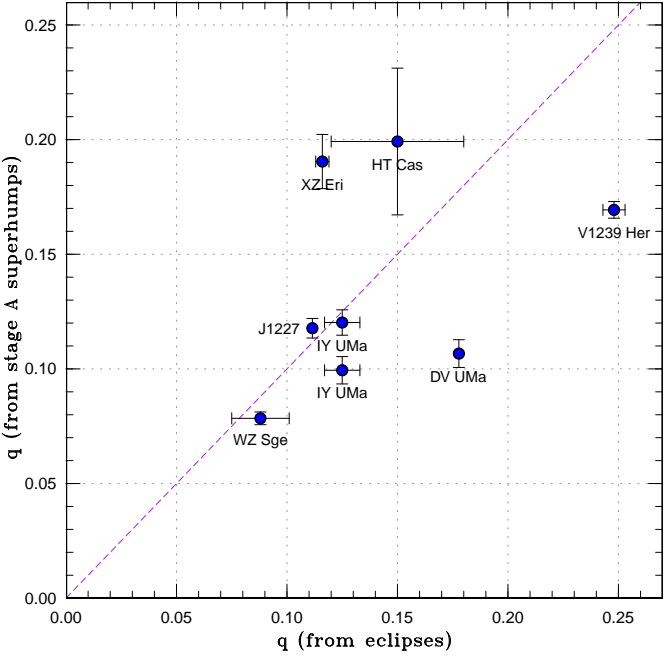}
  \end{center}
  \caption{Comparison of $q$ measured from eclipse (abscissa) and $q$
  estimated from stage A superhump based on our method (ordinate). 
  The names of the objects are given with labels
  (J1227 stands for SDSS J122740.83$+$513925.0)
  IY UMa has two observations (2000 and 2009), which are plotted
  individually.  The values of measured $q$ for all objects other than 
  WZ Sge were derived from eclipse observations, while estimated $q$ are 
  those obtained from superhump periods at the stage A by our method.}
  \label{fig:stageacomp}
\end{figure}

\begin{table*}
\caption{Comparison of $q$ directly measured from eclipse and $q$ estimated 
from stage A superhumps by our method.}\label{tab:stageawithq}
\begin{center}
\begin{tabular}{cccccc}
\hline
Object    & $P_{\rm orb}$\commenta\commentb & $P_{\rm SH}$ (stage A)\commenta 
& $q$(from eclipse)  & $q$ (by our method) & References\commentc \\
\hline
WZ Sge     & 0.056688 & 0.058385(56) & 0.092(8)\commentd & 0.078(3) & 1, 2, 3 \\
SDSS J1227 & 0.062950 & 0.065680(89) & 0.1115(16) & 0.118(4) & 4, 2, 5 \\
XZ Eri     & 0.061159 & 0.06519(21) & 0.116(3) & 0.190(12) & 6, 2, 6 \\
IY UMa     & 0.073909 & 0.07666(15) & 0.125(8) & 0.099(6) & 7, 2, 7 \\
           &          & 0.07718(14) &          & 0.120(6) & 8 \\
HT Cas     & 0.073647 & 0.07868(69) & 0.15(4) & 0.20(3) & 9, 9, 10 \\
DV UMa     & 0.085853 & 0.08926(18) & 0.1778(22) & 0.107(6) & 6, 2, 5 \\
V1239 Her  & 0.100082 & 0.10605(11) & 0.248(5) & 0.169(4) & 11, 12, 5 \\
\hline
\multicolumn{6}{l}{\commenta Unit d.} \\
\multicolumn{6}{l}{\commentb The error is smaller than 1 in the last
  significant digit.} \\
\multicolumn{6}{c}{\parbox{0.87\textwidth}{
\commentc The three numbers refer to the references to
$P_{\rm orb}$, $P_{\rm SH}$ (stage A) and $q$.
1: \citet{pat02wzsge}, updated $P_{\rm orb}$=0.0566878474~d was
used in \citet{Pdot};
2: \citet{Pdot};
3: \citet{ste07wzsge};
4: \citet{lit06j1702};
5: \citet{sav11CVeclmass};
6: \citet{fel04xzeridvuma};
7: \citet{ste03iyumaSTJ};
8: \citet{Pdot2};
9: \citet{Pdot3};
10: \citet{hor91htcas};
11: \citet{lit06j1702};
12: \citet{Pdot4};
}} \\
\multicolumn{6}{l}{\commentd From Doppler tomography.}
\end{tabular}
\end{center}
\end{table*}

   Some objects have known mass ratios determined from quiescent
eclipse observations.  These objects are listed in
table \ref{tab:stageawithq}
together with WZ Sge, which has constraint on $q$ from Doppler
tomography \citep{ste01wzsgesecondary}.
We have determined the $q$ from stage A superhumps
by our method described in section \ref{sec:model}, 
and made a comparison with the observed $q$.

   As the first step, we made a comparison of the estimated 
$q$ from all known stage A superhumps by our method and 
$q$ from quiescent eclipse observations to see what quality 
of observation is needed to make this estimate.
The result is shown in figure \ref{fig:stageacomp}.
Since there are intrinsic difficulties
both in eclipse observations and observation of stage A superhumps,
there remain a relatively large scatter in some objects.
The agreement is, however, good for objects with both reliable 
eclipse observations and well-covered stage A observation.

   The problem for each object can be summarized as follows. \\
{\it V1239 Her:} There were only three superhump measurements in stage A,
and the period was determined assuming that stage A continued to
$E=22$ \citep{Pdot4}.  If stage A--B transition took place earlier,
the period of stage A should be longer.  The present analysis
would favor this interpretation.  This period appears
to be inadequate for a comparison. \\
{\it DV UMa:} The eclipses of this object are deep, and it was
difficult to determine the times of stage A superhumps.
The combined $O-C$ diagram showed a large scatter \citep{Pdot3}.
Although the 2007 data were used to make the present comparison,
possible stage A superhumps were also detected in the 1997 data
(\cite{pat00dvuma}; \cite{Pdot}).  Using $E \le 12$ maxima for
the 1997 data, we obtained a period of 0.0933(9)~d.  This period
gave $q$=0.26, which appears too large.  The period of stage A
superhumps in this object needs to be re-examined by future
observations. \\
{\it HT Cas:} The measurement by \citet{hor91htcas} was rather old,
and the weakness of the hot spot in this object would make
the eclipse analysis difficult [e.g. \citet{fel05gycncircomhtcas}].
\citet{ioa99htcas} favored a different $q$ value.  It appears that
$q$ value is still uncertain for this object.
Despite the very good coverage of the 2010 superoutburst \citep{Pdot3},
the eclipses unfortunately overlapped the superhump maxima
around this phase (see figure 9 in \cite{Pdot3}) and the period
of stage A superhumps was very inaccurate.  This period appears
to be inadequate for a comparison. \\
{\it XZ Eri:} The 2008 data were used.  There were only five
measurements of superhump maxima with relatively large errors
\citep{Pdot}.  A combined $O-C$ diagram with the 2007 observation
(figure 87 in \cite{Pdot}) appears to suggest a shorter period
for stage A superhumps.  The period of stage A superhumps in this 
object again needs to be re-examined by future observations.

   Considering these uncertainties and difficulties, direct
comparisons of $q$ values from eclipse observations and
stage A superhumps will continue to be the challenges of
the future.  We can, however, safely say there does
not appear to be a strongly contradicting case at least
for low-$q$ objects.

\subsection{Evaluation of $q$ from Stage A Superhump and 
$P_{\rm orb}-q$ Relation}

   Although a comparison between $q$ values from stage A superhumps
and from quiescent eclipse observation is most direct, 
it suffers from the small number of objects and lower
quality of stage A observation due to eclipses.
We therefore took an alternative approach in which we examine 
the relation between the binary orbital period $P_{\rm orb}$ and 
the mass ratio $q$ estimated from stage A superhumps 
by our method for a much larger sample of both eclipsing and 
non-eclipsing SU UMa stars. 

   From evolutionary consideration of the cataclysmic variable stars, 
it is well known (e.g. \cite{kni06CVsecondary}) that there 
is a definite relation between the orbital period, $P_{\rm orb}$, 
and the mass of the secondary star ($P_{\rm orb}-M_2$ relation)
and if we assume some certain mass for the primary white dwarf 
[$M_1$; \citet{sav11CVeclmass} showed that the mean 
$M_1$=0.83 $M_\odot$ with an intrinsic scatter of 0.07 $M_\odot$
for systems with $P_{\rm orb} \le 0.066$~d.], 
we can translate it to $P_{\rm orb}-q$ relation. 

   In this subsection we examine the binary mass ratios $q$ estimated 
from the stage A superhump by using the $P_{\rm orb}-q$ relation.
We also compare the same relation for the mass ratio obtained 
from quiescent eclipse observation.

\subsubsection{Estimation of $q$ Values from Stage A Superhumps by 
  Our Method for a Much Larger Sample of Both Eclipsing and 
  Non-Eclipsing SU UMa Stars}

\begin{figure}
  \begin{center}
    \FigureFile(88mm,70mm){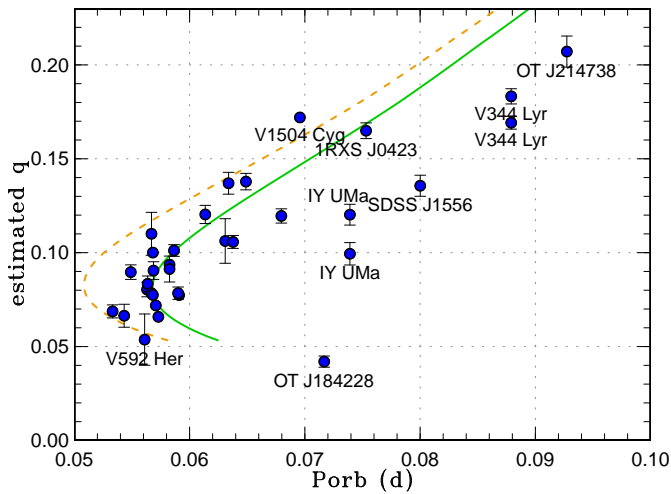}
  \end{center}
  \caption{Estimated $q$ based on our method versus $P_{\rm orb}$.
  Names are shown for only some objects.
  The dashed and solid curves represent the standard and optimal
  evolutionary tracks in \citet{kni11CVdonor}, respectively.
  }
  \label{fig:qfromstagea}
\end{figure}

   We estimated $q$ from periods of stage A superhumps by our method 
for a much large sample of both eclipsing and non-eclipsing 
SU UMa stars.   Since it has become evident (subsection
\ref{sec:knownq}) that not all values for stage A superhump 
listed in the study like \citet{Pdot} are adequate 
for this purpose, we restricted our analysis to observations in which
stage A--B transition was recorded (table \ref{tab:qestfromstagea}).
The quality 1 represents multiple-night observation of stage A
and quality 2 represents either single-night observation or
observations with relatively large errors.
For these objects, the difficulty met in deeply eclipsing systems
(subsection \ref{sec:knownq}) that eclipses interfere 
the detection and measurement of stage A superhumps 
does not exist, and we only need to focus on
the duration and selection of the segment used for analysis.
Based on necessary requirement for observation as examined in
subsection \ref{sec:knownq}, we restricted $E$ ranges for
the three objects: $E \le 13$ for SDSS J1556 (2007),
$E \le 17$ for PU CMa (2008) and $8 \le E \le 21$ for
BW Scl (2011, due to the contamination of early superhumps)
obtained revised values.  The random error in period estimates
with the PDM and $O-C$ analysis by the profile fitting method
\citep{Pdot} sufficiently lower than 0.0002~d if the observation
baseline is 1--2~d, which typically corresponds to an error
of 0.01 in $q$.  In low-$q$ systems, the stage A lasts longer
and its period is easier to determine.

   We also noticed that there was a better segment in
the Kepler data of V1504 Cyg to measure stage A superhumps
(Osaki and Kato in prep.).  In this case, stage A superhumps were 
observed during the quiescent state between the precursor and 
the main superoutburst.
Since the precession frequency is not expected to be affected
by the pressure effect in such a cold disk, we can safely neglect
the pressure effect.  Using the V1504 Cyg data for
BJD 2455876.8--2455878.7, we have obtained a period of
stage A superhump 0.07377(6)~d.  This corresponds to
$\epsilon^*$=0.0570(7) and the value of $q$=0.172(2)
is derived.  These values were adopted in table
\ref{tab:qestfromstagea}.  A closer examination of the $O-C$ data 
in \citet{Pdot3} indicates that the period of stage A superhumps
was variable even within stage A, and the period corresponding
to 0.07377~d was only present on the first night of
the appearance of stage A.  We interpret that the tidal effect
is stronger in higher-$q$ systems and the eccentric region
spreads more quickly than in low-$q$ systems.
This lesson would tell us we should pay more attention to
choosing a segment of observations when analyzing objects 
with long-$P_{\rm orb}$ or higher-$q$ objects.
The value of 0.07395(21)~d in \citet{Pdot3} corresponds to
$q$=0.181(10).  The revised value is within 1$\sigma$ of
the value derived from the entire stage A superhumps 
in an ordinary superoutburst, and these measurements
are not in serious contradiction.

   The result is shown in figure \ref{fig:qfromstagea} for 
the $P_{\rm orb}-q$ relation.  
In table \ref{tab:qestfromstagea}, we also listed $q$ values
in \citet{pat11CVdistance} (P11).  Most of his $q$ values 
were determined from $\varepsilon$ for stage B superhumps using 
the empirical relation in \citet{pat05SH}.  For objects with 
high-quality stage A measurements, the agreement between these 
two methods is very good for $q$ larger than
0.085--0.090, indicating that our method is consistent with
the previous method at least for higher-$q$ objects.
\citet{pat05SH} used an older $P_{\rm orb}$ for V342 Cam
[0.0763~d, one of the candidate aliases listed in \citet{aun06HSCV},
see \citet{Pdot} for the selection], and the difference
between the estimated $q$ values was caused by this period
selection.  If we use the older value, we obtain $q$=0.121(4),
perfectly in agreement.

   The agreement is, however, worse for lower $q$ objects.
This is particularly clearly seen in WZ Sge, for which
\citet{pat11CVdistance} gave $q$=0.046 in contrast to
our $q$=0.078(3).  \citet{pat11CVdistance} seems to have
underestimated $q$ for low-$q$ objects, i.e. WZ Sge-type
dwarf novae.  This can be understood as follows: 
\citet{pat11CVdistance} used stage B superhumps, which is 
strongly affected by the pressure effect and the relative 
strength of the pressure effect depends on $q$, 
i.e. lower $q$ objects has a stronger pressure effect 
(\cite{kat13j1939v585lyrv516lyr}; later discussion in
this paper).  Since \citet{pat11CVdistance} used a nearly
linear $\varepsilon-q$ relation for low-$q$ and assumed
$\varepsilon=0$ for $q=0$, this pressure
effect was translated to a systematically smaller $q$.
This systematic trend would affect the discussion
for the evolution of low-$q$ objects, i.e. the objects
near or after the period minimum.  For example, systematically
smaller $q$ will increase the fraction of brown-dwarf
secondaries or period bouncers.  Our figure \ref{fig:qfromstagea}
suggests that most of the dwarf novae have not yet reached
the period minimum.

   One notable object is OT J184228.  This object was suggested
to be a period bouncer using the $\varepsilon$ for stage B superhumps
(which were only seen during its second outburst; \cite{Pdot4}).
The present conclusion from stage A superhumps well agrees with
the previous identification.

\begin{table*}
\caption{$q$ estimated from stage A superhumps by our method.}
 \label{tab:qestfromstagea}
\begin{center}
\begin{tabular}{cccccccc}
\hline
Object & Year & $P_{\rm orb}$\commenta\commentb & $P_{\rm SH}$\commenta (stage A) & $q$ (by our method) & $q$ (from  P11) \commentc & Quality\commentd & References\commente \\
\hline
\citet{Pdot} \\
V455 And & 2007 & 0.05631 & 0.05803(8) & 0.080(4) & 0.06 & 1 & 1, 2 \\
V466 And & 2008 & 0.05636 & 0.05815(8) & 0.083(4) & 0.058 & 1 & 2, 2 \\
VY Aqr & 2008 & 0.06309(4) & 0.06558(26) & 0.106(12) & 0.095 & 1 & 3, 2 \\
V342 Cam & 2008 & 0.07531(8) & 0.07970(10) & 0.164(4) & 0.121 & 1 & 4, 2 \\
WX Cet & 1989 & 0.05826 & 0.06031(3) & 0.094(1) & 0.094 & 1 & 5, 2 \\
       & 1998 &         & 0.06027(14) & 0.091(7) &  & 1 & 2 \\
HO Cet & 2006 & 0.05490(2) & 0.05676(7) & 0.090(4) & 0.091 & 1 & 2, 2 \\
PU CMa & 2008 & 0.05669(4) & 0.05901(22) & 0.110(11) & 0.109 & 2 & 6, 2 \\
V632 Cyg & 2008 & 0.06377(8) & 0.06628(7) & 0.106(3) & 0.125 & 2 & 7, 2 \\
GW Lib & 2007 & 0.05332(2) & 0.05473(7) & 0.069(3) & 0.056 & 1 & 8, 2 \\
V453 Nor & 2005 & 0.06338(4) & 0.06653(12) & 0.137(6) & 0.082 & 1 & 9, 2 \\
UV Per & 2003 & 0.06489(1) & 0.06813(9) & 0.138(4) & 0.108 & 1 & 3, 2 \\
V493 Ser & 2007 & 0.08001(1) & 0.08395(14) & 0.136(6) & 0.156 & 2 & 10, 2 \\
WZ Sge & 2001 & 0.05669 & 0.05838(6) & 0.078(3) & 0.046 & 1 & 11, 2 \\
SW UMa & 2006 & 0.05681 & 0.05894(5) & 0.100(3) & 0.113 & 1 & 12, 2 \\
IY UMa & 2000 & 0.07391 & 0.07666(15) & 0.099(6) & 0.12 & 1 & 13, 2 \\
KS UMa & 2003 & 0.06796(10) & 0.07095(8) & 0.120(4) & 0.112 & 1 & 14, 2 \\
HV Vir & 2002 & 0.05707 & 0.05864(2) & 0.072(1) & 0.094 & 2 & 14, 2 \\
ASAS J1025 & 2006 & 0.06136(6) & 0.06407(10) & 0.120(5) & 0.135 & 1 & 2, 2 \\
\hline
\citet{Pdot2} \\
V592 Her & 2010 & 0.05610\commentf & 0.05728(29) & 0.054(14) & 0.037? & 2 & 15, 15 \\
IY UMa & 2009 & 0.07391 & 0.07717(14) & 0.120(6) & 0.12 & 1 & 12, 15 \\
SDSS J1610 & 2009 & 0.05687(1) & 0.05881(9) & 0.090(5) & 0.086 & 1 & 15, 15 \\
OT J1044 & 2010 & 0.05909(1) & 0.06084(3) & 0.077(1) & -- & 2 & 15, 15 \\
\hline
\citet{Pdot3} \\
EZ Lyn & 2010 & 0.05901 & 0.06077(7) & 0.078(3) & 0.05 & 1 & 16, 16 \\
V344 Lyr & 2009 & 0.08790 & 0.09314(9) & 0.169(3) & -- & 1 & 17, 16 \\
V344 Lyr & 2009b & 0.08790 & 0.09351(10) & 0.183(4) & -- & 1 & 17, 16 \\
SW UMa & 2010 & 0.05681 & 0.05850(2) & 0.077(1) & 0.113 & 2 & 11, 16 \\
V355 UMa & 2011 & 0.05729 & 0.05874(3) & 0.066(1) & -- & 1 & 18, 16 \\
\hline
\citet{Pdot4} \\
BW Scl & 2011 & 0.05432 & 0.05572(12) & 0.067(6) & -- & 1 & 19, 20 \\
OT J184228 & 2011 & 0.07168(1) & 0.07287(8) & 0.042(3) & -- & 1 & 20, 20 \\
OT J210950 & 2011 & 0.05865\commentf & 0.06087(6) & 0.101(3) & -- & 2 & 20, 20 \\
OT J214738 & 2011 & 0.09273\commentf & 0.09928(22) & 0.207(8) & -- & 1 & 20, 20 \\
\hline
This paper \\
V1504 Cyg & 2009b & 0.06955 & 0.07377(6) & 0.172(2) & 0.150 & 1 & 16, 21 \\
\hline
\multicolumn{8}{l}{\commenta Unit d.} \\
\multicolumn{8}{l}{\commentb The error is smaller than 1 in the last
  significant digit if the error is omitted.} \\
\multicolumn{8}{l}{\commentc $q$ value in \citet{pat11CVdistance}.
  Most of the values were determined from $\varepsilon$ for stage B
  superhumps.} \\
\multicolumn{8}{l}{\commentd Quality of stage A observation.  1: good, 2: low quality.} \\
\multicolumn{8}{c}{\parbox{1.04\textwidth}{
\commente The two numbers refer to the references to
$P_{\rm orb}$ and $P_{\rm SH}$ (stage A).
1: \citet{ara05v455and};
2: \citet{Pdot};
3: \citet{tho97uvpervyaqrv1504cyg};
4: \citet{she11j0423};
5: \citet{ste07wxcet};
6: \citet{tho03kxaqlftcampucmav660herdmlyr};
7: \citet{she07CVspec};
8: \citet{tho02gwlibv844herdiuma};
9: \citet{ima06asas1600};
10: \citet{wou04CV4}, alias selection in \citet{Pdot};
11: \citet{pat02wzsge}, \citet{Pdot};
12: J. Thorstensen, PhD thesis
13: \citet{ste03iyumaSTJ};
14: \citet{pat03suumas};
15: \citet{Pdot2};
16: \citet{Pdot3};
17: \citet{osa13v344lyrv1504cyg};
18: \citet{gan06j1339}, updated in \citet{Pdot3};
19: \citet{aug97bwscl};
20: \citet{Pdot4}.
21: this paper.
}} \\
\multicolumn{8}{l}{\commentf Tentative identification.} \\
\multicolumn{8}{c}{\parbox{1.04\textwidth}{
Abbreviations for the names: ASAS J102522$-$1542.4 (ASAS J1025),
SDSS J161027.61$+$090738.4 (SDSS J1610),
OT J104411.4$+$211307 = CSS100217:104411$+$211307 (OT J1044), 
OT J184228.1$+$483742 (OT J184228), 
OT J210950.5$+$134840 (OT J210950), 
OT J214738.4$+$244553 = CSS111004:214738$+$24455 (OT J214738).}} \\
\end{tabular}
\end{center}
\end{table*}

\subsubsection{Comparison of $q$ Values with Those from Quiescent
Eclipse Observation}\label{sec:eclqcomp}

   In recent years, the advent of high-speed CCD photometry
and the use of Markov-chain Monte Carlo (MCMC) modeling
of the light curve have made great improvement in measuring
binary parameters (\cite{lit08eclCV}; \cite{sou11j1808})
than in the 1990s.  We can now use more reliable set of
$q$ measurements than in the time of \citet{pat05SH},
\citet{kni06CVsecondary}.

\begin{figure}
  \begin{center}
    \FigureFile(88mm,70mm){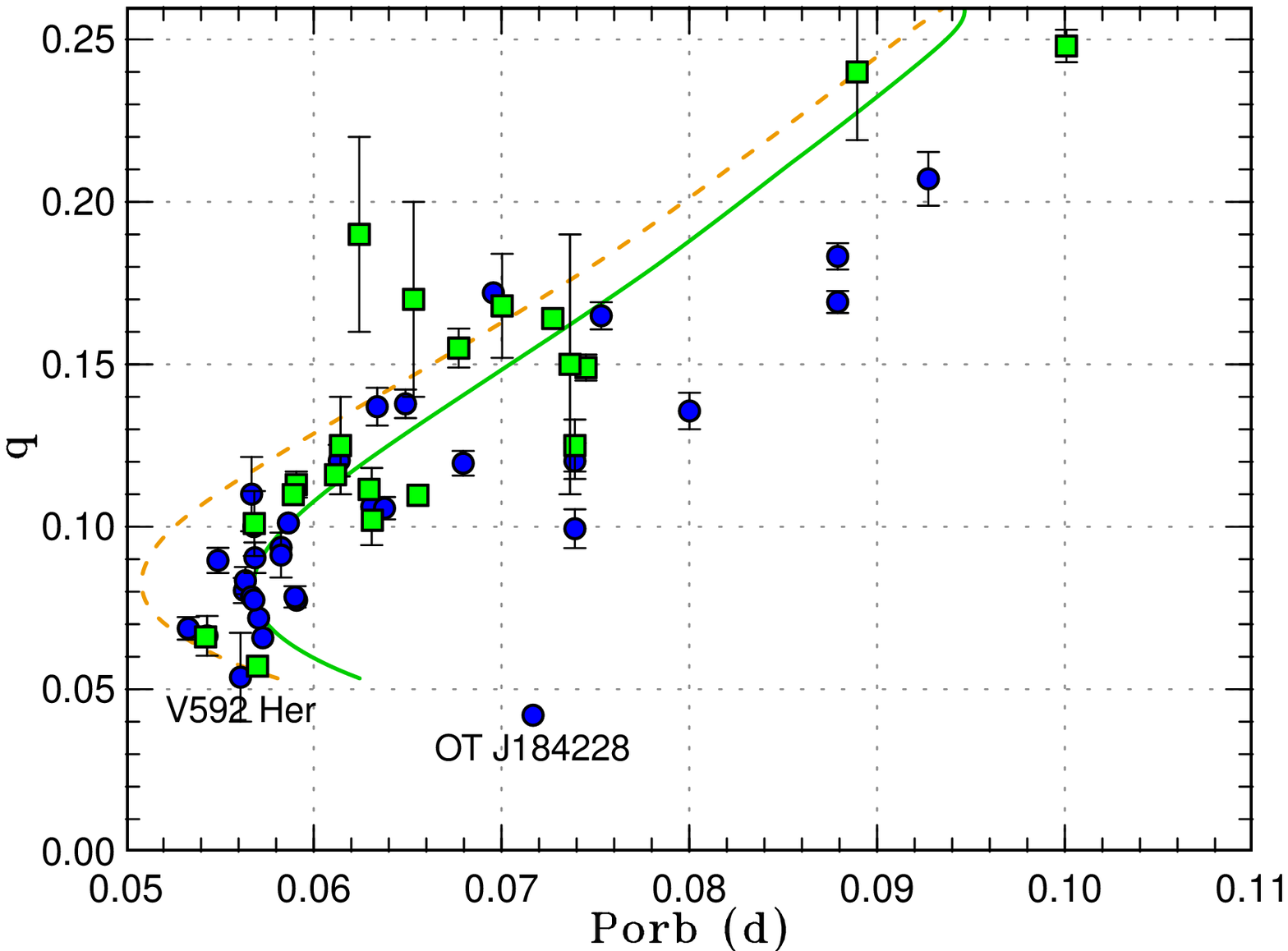}
  \end{center}
  \caption{Comparison of $q$ estimated by our method from stage A superhumps
  (filled circles) and $q$ measured by eclipses (filled squares).
  The agreement is very good.
  The dashed and solid curves represent the standard and optimal
  evolutionary tracks in \citet{kni11CVdonor}, respectively.
  Our new $q$ values support the discussion on the evolutionary sequence 
  in \citet{lit08eclCV} and \citet{kni11CVdonor} in that
  an angular momentum loss larger than solely from the gravitational
  wave radiation is needed to reproduce the $P_{\rm orb}$-$q$ relation.}
  \label{fig:qall}
\end{figure}

   A comparison of $q$ estimated from stage A superhumps
and $q$ measured by eclipses is given in figure \ref{fig:qall}.
In addition to the objects in table \ref{tab:stageawithq},
we included Z Cha \citep{wad88zcha}, OY Car \citep{lit08eclCV},
1RXS J180834.7$+$101041, V1258 Cen, CTCV J2354$-$4700,
SDSS J115207.00$+$404947.8, OU Vir, SDSS J103533.02$+$055158.3,
SDSS J090350.73$+$330036.1, SDSS J143317.78$+$101123.3,
NZ Boo, SDSS J150137.22$+$550123.4 \citep{sou11j1808},
V2051 Oph \citep{bap98v2051ophHST}, SDSS J152419.33$+$220920.0
\citep{sou10SDSSeclCV} and V4140 Sgr \citep{bor05v4140sgr}.
The agreement of the evolutionary sequence between two
types of measurements is very good.  The outlier in the upper
left corner is V2051 Oph, whose quality of parameter estimation
may have not been so good.  This figure corresponds to figure 6
in \citet{lit08eclCV} with a much higher number of $q$ estimates.

   We have seen both in direct comparison and in comparison of
the inferred evolutionary sequence that $q$ values from stage A
superhumps are reliable and as accurate as those from determined
from quiescent eclipse observations.  We can therefore
adopt the assumption that stage A superhumps indeed reflect
the dynamical precession rate at the radius of the 3:1 resonance.
We should note, however,
the result by our method appears to give smaller $q$ for
long-$P_{\rm orb}$ objects.  We suspect that this is caused
by insufficient observational coverage for early part of
stage A superhumps in these objects, since these objects took
shorter time to develop superhumps, and the early stage A
evolution tends to be missed by observation.  This difficulty
is apparently avoided for short-$P_{\rm orb}$ objects because
they have a long waiting time before superhumps appears.
This interpretation need to be checked by future observations.

\section{Discussion}\label{sec:discussion}

\subsection{New Calibration for $\varepsilon-q$ Relation}

   Although we consider that the use of stage A superhumps 
is the first choice for estimating $q$ for objects with
well-observed stage A superhumps, this condition is not
always met by observation.  In such cases, we may use
stage B superhumps to estimate $q$, as have been done traditionally,
under a simplified assumption of the constant pressure effect
for each $q$.  One should remember that the pressure effect
may different between different objects, and the treatment
here may not be always valid.

\begin{figure}
  \begin{center}
    \FigureFile(88mm,70mm){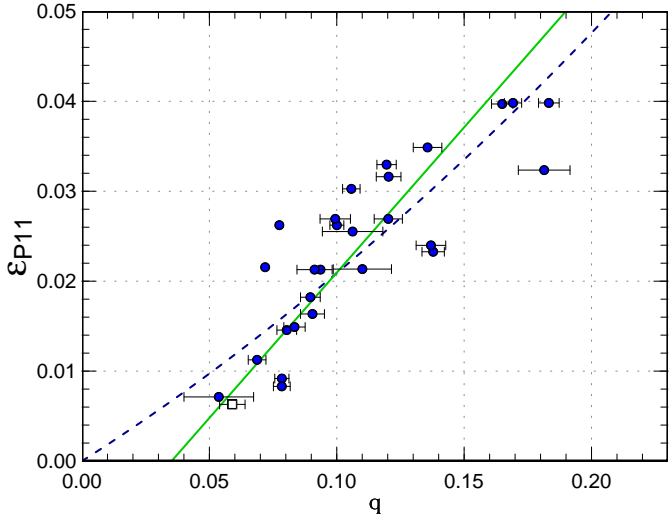}
  \end{center}
  \caption{Relation between $\varepsilon$ \citep{pat11CVdistance}
  and $q$ estimated from stage A superhumps.
  The dashed curve and solid line represent the relation in
  \citet{pat05SH} and a linear fit, respectively.
  The open rectangle represents EG Cnc, estimated from post-superoutburst
  superhumps (see subsection \ref{sec:diskradiuspostSO}), which was not
  used for the regression. 
  }
  \label{fig:qepspat}
\end{figure}

   By using $q$ estimated from stage A superhumps, we can re-calibrate
the $\varepsilon-q$ relation for a much larger sample than in
\citet{pat05SH} or \citet{Pdot}, which were mainly based on $q$ values 
estimated by using relatively rare eclipsing systems.
As the first step, we re-calibrate the frequently used $\varepsilon-q$ 
relation in \citet{pat05SH} (and its improvement in \cite{Pdot}).
In order to make a fair comparison with \citet{pat05SH},
we used $\varepsilon$ estimated from $P_{\rm SH}$ tabulated
in his paper by \citet{pat11CVdistance} since $\varepsilon$
is dependent on the outburst phase, and \citet{pat05SH} apparently
``estimated period 4~d after maximum light
(or hump onset)'' when this is available [footnote 11 of
\citet{pat11CVdistance}].  Since $P_{\rm SH}$ for WZ Sge was missing,
we supplied it from \citet{pat02wzsge}.  The result is shown
in figure \ref{fig:qepspat}.  The systematic departure from
the relation in \citet{pat05SH} is evident: $q$ is always
estimated lower for $\varepsilon \le 0.018$.
A linear fit yielded a new relation
\begin{equation}
\label{equ:qpat11}
q=0.035(10)+3.09(40)\varepsilon ,
\end{equation}
which may be used for superhump periods in the early phase
(but not so early as stage A superhumps).
This equation suggests that $\varepsilon$ could be negative for $q\le 0.035$ 
(i.e., retrograde precession for a sufficiently low $q$),  
and this is probably due to the stronger pressure effect
in relation to the tidal effect in low-$q$ systems.

   Although this functional form is not theoretically derived,
we may have a qualitative interpretation.  According to
\citet{lub92SH}, the apsidal precession rate ($\nu_{\rm pr}$)
can be written as a form:
\begin{equation}
\nu_{\rm pr}=\nu_{\rm dyn}+\nu_{\rm pressure}+\nu_{\rm stress},
\label{equ:Lubows}
\end{equation}
where the first term, $\nu_{\rm dyn}$, represents a contribution 
to disk precession due to tidal perturbing force of the secondary, 
giving rise to prograde precession, the second term, $\nu_{\rm pressure}$, 
the pressure effect giving rise to retrograde precession, and 
the last term, $\nu_{\rm stress}$, the minor wave-wave interaction.
While $\nu_{\rm dyn}$ is dependent on $q$ (roughly $\propto q$),
$\nu_{\rm pressure}$ is independent of $q$, suggesting
a functional form like $\varepsilon \sim a \times q - {\rm const.}$, a linear 
relation between $\varepsilon$ and $q$.  Although this effect was also
noticed in the analyses by \citet{pea06SH} and \citet{goo06SH},
their samples were a mixed class of objects from \citet{pat05SH}.
The actual treatment of the precession rate in the presence of
pressure, however, is more complex in terms of eigenfunction expression
(cf. \cite{hir93SHperiod}; \cite{osa13v344lyrv1504cyg}).

   Similarly, we obtained a relation for the middle (mean)
of stage B superhumps (values from \cite{Pdot}, \cite{Pdot2},
\cite{Pdot3}, \cite{Pdot4}) as follows:
\begin{equation}
\label{equ:stagebmid}
q=0.026(7)+3.36(27)\varepsilon .
\end{equation}

\begin{figure}
  \begin{center}
    \FigureFile(88mm,70mm){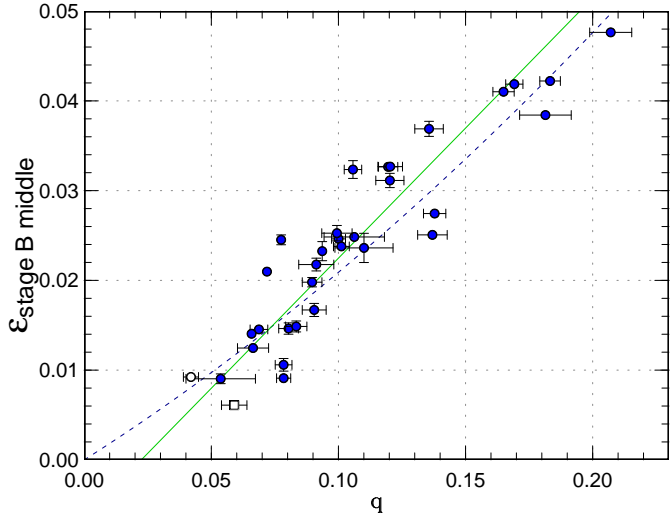}
  \end{center}
  \caption{Relation between $\varepsilon$ for the middle (mean)
  of stage B superhumps and $q$ estimated from stage A superhumps.
  The dashed curve and solid line represent the relation in
  \citet{pat05SH} and a linear fit, respectively.
  The open circle represents the unusual object OT J184228.
  The open rectangle represents EG Cnc, estimated from post-superoutburst
  superhumps (see section \ref{sec:diskradiuspostSO}).
  Both objects were not used for the regression.}
  \label{fig:qepsb2}
\end{figure}

\begin{figure}
  \begin{center}
    \FigureFile(88mm,70mm){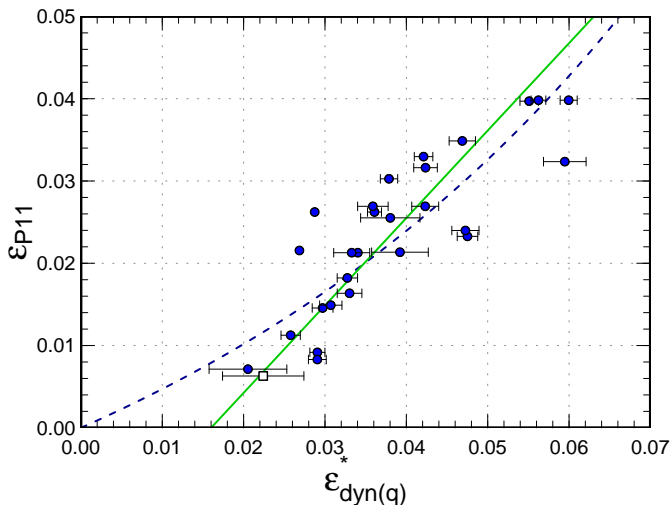}
  \end{center}
  \caption{Relation between $\varepsilon$ \citep{pat11CVdistance}
  and $\varepsilon^*$ stage A superhumps which represents
  $q$.  The dashed curve and solid line represent the relation in
  \citet{pat05SH} and a linear fit, respectively.
  The open rectangle represents EG Cnc, estimated from post-superoutburst
  superhumps (see subsection \ref{sec:diskradiuspostSO}), which was not
  used for the regression. 
  }
  \label{fig:dynepspat}
\end{figure}

   Although these regressions appear useful within the limit
of observational scatter, a linear regression on the $q-\varepsilon$
space cannot properly take into account of 
the non-linear dependence of  $\varepsilon$ on $q$
(we therefore do not recommend to use figures
\ref{fig:qepspat}, \ref{fig:qepsb2} to estimate $q$).
We instead may better use the value of $\varepsilon^*$ 
at the 3:1 resonance radius
(i.e.,  the value of  $\varepsilon^*$ for stage A superhumps) 
to represent $q$ and we find a relation between observed $\varepsilon$ 
and this dynamical precession rate $\varepsilon^*$ in representing $q$. 
This method is expected to better deal with the non-linear
functional form of $\varepsilon^*$ against $q$.
Once this relation is obtained, one can estimate the dynamical 
precession rate for the 3:1 resonance radius from the observed 
$\varepsilon$, and then convert the dynamical precession rate into 
$q$ using the relation in section \ref{sec:model} (such as table 
\ref{tab:stageaepsq} or figure \ref{fig:qeps31}). 
The result is shown in figure \ref{fig:dynepspat}. 
The presence of a curvature of the relation
by \citet{pat05SH} on this plot now clearly demonstrates
the cause of the systematic errors for low-$q$ objects using
the relation in \citet{pat05SH}.  The linear regression is
\begin{equation}
\label{equ:dynepspat}
\varepsilon^*(3:1)=0.016(3)+0.94(12)\varepsilon .
\end{equation}
We recommend to use this relation rather than equations
(\ref{equ:qpat11}) and (\ref{equ:stagebmid}) to estimate $q$ 
from early stage (early stage B) superhump observations.

   Figure \ref{equ:dynepsb2} is the same relation for the middle (mean)
of stage B superhumps.  The relation is
\begin{equation}
\label{equ:dynepsb2}
\varepsilon^*(3:1)=0.012(2)+1.04(8)\varepsilon .
\end{equation}

\begin{figure}
  \begin{center}
    \FigureFile(88mm,70mm){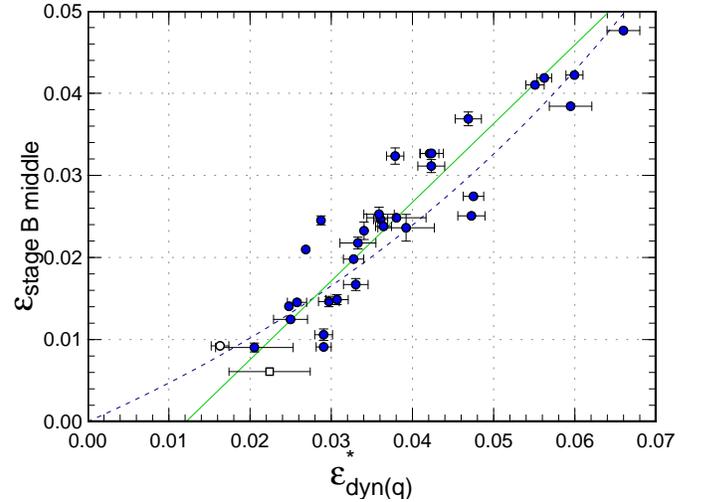}
  \end{center}
  \caption{Relation between $\varepsilon$ for the middle (mean)
  of stage B superhumps and $\varepsilon^*$ stage A superhumps which 
  represents $q$.  The dashed curve and solid line represent 
  the relation in \citet{pat05SH} and a linear fit, respectively.
  The open rectangle represents EG Cnc, estimated from post-superoutburst
  superhumps (see subsection \ref{sec:diskradiuspostSO}), which was not
  used for the regression. 
  }
  \label{fig:dynepsb2}
\end{figure}

\subsection{Implication to CV Evolution}\label{sec:evolimpl}

   The new $q$ values estimated from stage A superhumps make
the largest difference in low-$q$ or short-$P_{\rm orb}$ systems,
i.e. WZ Sge-type dwarf novae.  Since the case of WZ Sge itself has
been long discussed, here we examine this case.
Despite a huge amount of efforts, the nature of the secondary in
WZ Sge is still unclear.  \citet{ste01wzsgesecondary} first
succeeded in detecting the emission from the secondary in 
the Doppler tomography, and \citet{ste07wzsge} combined the
ultraviolet radial velocity to estimate $M_1$=0.85(4)$M_{\odot}$
and $M_2$=0.078(6)$M_{\odot}$.  \citet{pat02wzsge}, however,
argued that $q$ should be smaller considering the $M_K \sim 9.5$
for a 0.09$M_{\odot}$ secondary which contradict with $M_K > 12.2$
derived from a trigonometric distance of 43(2) pc
\citep{tho03CVdistance} and the absence of the secondary feature
in the spectrum.  \citet{pat11CVdistance} preferred a very
low mass secondary, $q=0.045$.

   Our result suggests $q$=0.078(3), and using \citet{ste07wzsge},
we obtained $M_2$=0.066(4)$M_{\odot}$.
This mass corresponds to a brown dwarf
having $M_K$=10.6--11.1 \citep{kni06CVsecondary}.
\citet{har13CVHerschel} recorded the near-/mid-infrared variation
attributable to ellipsoidal modulations of an L2-type secondary
having $M_K$=11.1.  \citet{har13CVHerschel} estimated that
the secondary contributes to 80\% at the eclipse minimum,
and raised a question why the expected CO absorption for an L2
brown dwarf has not yet been detected.  Our mass estimate higher
than \citet{pat02wzsge} supports the result by \citet{har13CVHerschel}
and the secondary of WZ Sge is likely an object on the borderline 
of the end of the main sequence ($M_K \sim 11$) and 
brown dwarfs.\footnote{
   In reality, the popular Kumar limit ($\sim$0.075
   $M_{\odot}$) for core hydrogen burning has no great significance
   in the case of CV evolution.  The surface temperature of
   the secondary can depend on the age and evolution history.
   We use the term ``brown dwarf'' in the popular sense
   when describing the CV secondaries, and mean secondaries
   with masses somewhere below 0.05--0.09 $M_{\odot}$
   \citep{pat11CVdistance}.
}

   The $q$ value for WZ Sge is slightly higher than the value 
of $q$=0.0661(7) for SDSS J143317.78$+$101123.3, the eclipsing object 
with a system parameter ($P_{\rm orb}$=0.05424~d) close to WZ Sge.
The secondary of this object has recently been identified as
an L2-type brown dwarf \citep{lit13j1433}.  There have been
another detection of a brown dwarf in the eclipsing system
SDSS 103533.03$+$055158.4 having $P_{\rm orb}$=0.057007~d and
$q$=0.055(2) \citep{lit06j1035}.  Since the latter two objects
were more white dwarf-dominated in the spectrum and the eclipse
light curves, these objects can be naturally understood as
objects that passed the evolutionary stage of WZ Sge.

   There has been a lot of discussions regarding the value of
the period minimum and on the observational evidence for
its presence.  While the standard evolutionary model of CVs
gives a shorter period (65--70 min, \cite{kol99CVperiodminimum};
\cite{how01periodgap}), this value is much shorter than the minimum
orbital period of ordinary hydrogen-rich CVs \citep{kol93CVpopulation}.
\citet{gan09SDSSCVs} detected a high concentration of CVs near 
the period of $\sim$80 min using the sample selected by 
the Sloan Digital Sky Survey (SDSS).  \citet{wou12SDSSCRTSCVs}
used the sample based on the Catalina Real-Time Survey (CRTS)
and reached a similar conclusion.
These authors used the distribution of $P_{\rm orb}$
for discussion.  The $P_{\rm orb}$ distribution of dwarf novae 
detected as transients, however, can be biased due to the very low
outburst frequency near the period minimum.  Considering this bias,
\citet{uem10DNshortP} applied a Bayesian approach to estimate 
the parent population, and obtained a shorter period minimum
of $\sim$70 min assuming a certain functional form for
the $P_{\rm orb}$ distribution.  \citet{kat12DNSDSS} applied
the same method to the neural network-based $P_{\rm orb}$ estimates 
using the SDSS colors of CRTS-selected objects.  Although
the latter method has a potential to estimate $P_{\rm orb}$
for a large number of less biased transients, it still suffers 
from the spread of the distribution due to estimation errors,
and it allows a lower period minimum.  Our present method can provide
the $q$ distribution against $P_{\rm orb}$ even for faint objects,
and is more sensitive to the CV evolution than
the distribution of $P_{\rm orb}$ alone.  Our new result
indicates that $q$ starts to decrease quickly around
$P_{\rm orb} \simeq$80 min, and the evolutionary sequence
appears to approach the period minimum around this period.
Our result is not in contradiction with an assumption that 
the observed minimum period ($\sim$77 min) is indeed 
the period minimum.
This discussion, however, ignores the intrinsic spread 
of the period minimum and a more sophisticated treatment, 
as well as the increase of the sample including ones 
in the region of period bouncers, is required to make a more 
decisive discussion.

   In subsection \ref{sec:eclqcomp}, we suggested that the result 
by our method appears to give smaller $q$ for long-$P_{\rm orb}$ 
objects.  This is more apparent when we compare the result
with the theoretical evolutionary sequence (figures
\ref{fig:qfromstagea}, \ref{fig:qall}).  Since the Kepler data
for V344 Lyr, which was not affected by the lack of early stage A 
observations, deviates from the theoretical evolutionary
sequence, there may be a wider intrinsic spread of $q$ in
long-$P_{\rm orb}$ systems than in short-$P_{\rm orb}$ systems.
This possibility needs to be tested by a larger sample.

\subsection{Disk Radius in Post-superoutburst Stage}

\subsubsection{Disk Radius in WZ Sge-Type Dwarf Novae in Post-Superoutburst}
\label{sec:diskradiuspostSO}

   Another good application to the present method would be to estimate
the disk radius using superhumps long after the superoutburst.
Since the disk is sufficiently cold in such a situation, we can expect that
the superhump wave is confined to 
the disk's outer edge, and we can ignore the pressure effect.
Such persistent superhumps have been particularly well
observed in WZ Sge-type dwarf novae.  \citet{kat08wzsgelateSH}
estimated the radius by assuming that stage B superhump represents
the 3:1 resonance and concluded that these modulations are superhumps 
arising from matter near the tidal truncation radius.
Since the new assumption is very different from the one at
the time of \citet{kat08wzsgelateSH}, we re-estimated the radius.
The results are listed in table \ref{tab:rdiskfromlsh}.
It has become evident that there are two groups: objects with
a large disk radius (0.37--0.38$A$: GW Lib, V455 And, BW Scl) and
ones with a small disk radius (0.30--0.32$A$: WZ Sge, EZ Lyn).
These groups correspond to WZ Sge-type dwarf novae without
rebrightenings and with multiple rebrightenings, respectively.
It would suggest that the remnant disk matter is larger if
no rebrightenings occurs.  While the smaller one appears to
agree to what was supposed in \citet{osa95wzsge}, the larger one
is much larger than this.

\begin{table*}
\caption{Disk radii estimated from the periods of late-stage superhumps of 
WZ Sge-type dwarf novae.}\label{tab:rdiskfromlsh}
\begin{center}
\begin{tabular}{cccccc}
\hline
Object & $P_{\rm orb}$\commenta & $P_{\rm lsh}$\commentb
       & $q$ & $R_{\rm d}$\commentc & References \\
\hline
GW Lib   & 0.05332(2) & 0.054156(1) & 0.069 & 0.38 & \citet{kat08wzsgelateSH} \\
V455 And & 0.05630921(1) & 0.057280(4) & 0.080 & 0.37 & \citet{kat08wzsgelateSH} \\
WZ Sge   & 0.0566878460(3) & 0.057408(4) & 0.078 & 0.32 & \citet{kat08wzsgelateSH} \\
EZ Lyn   & 0.05900495(3) & 0.05967(2) & 0.078 & 0.30 & \citet{Pdot3} \\
V355 UMa & 0.057289(1) & 0.058183(5) & 0.066 & 0.38 & \citet{Pdot3} \\
BW Scl   & 0.0543234(7) & 0.055100(2) & 0.067 & 0.37 & \citet{Pdot4} \\
\hline
 \multicolumn{6}{l}{\commenta Orbital period (d).} \\
 \multicolumn{6}{l}{\commentb Period of late-stage superhumps (d).} \\
 \multicolumn{6}{l}{\commentc Estimated disk radius (unit in binary separation).} \\
\end{tabular}
\end{center}
\end{table*}

   If these radii for the WZ Sge-type dwarf novae in
the post-superoutburst state is universal, we may estimate $q$
by using the period of the post-superoutburst superhumps.
A notable example is EG Cnc \citep{pat98egcnc}, which showed
a period of 0.06051(2)~d (value refined in \cite{Pdot}).
EG Cnc is a system with multiple rebrightenings (\cite{pat98egcnc};
\cite{kat04egcnc}), and we assume a disk radius of 0.31$A$.
The resultant $q$=0.059(5), which is again larger than $q=0.035$
in \citet{pat11CVdistance}.  This value seems to be consistent
with the updated empirical $\varepsilon-q$ (figure \ref{fig:qepspat}).

\subsubsection{Disk Radius from Stage C Superhumps}
\label{sec:radiusfromstageC}

\begin{figure}
  \begin{center}
    \FigureFile(88mm,70mm){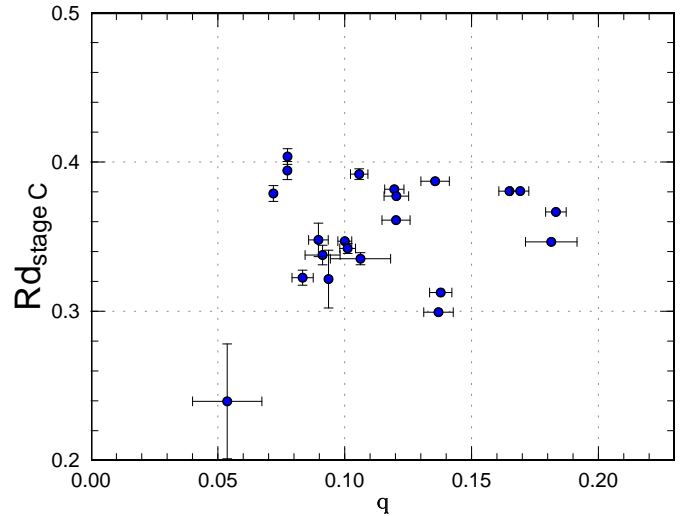}
  \end{center}
  \caption{Disk radius estimated from stage C superhumps, assuming
  that the pressure effect can be neglected.}
  \label{fig:qrstagec}
\end{figure}

   The same procedure may be applied to stage C superhumps,
though the disk is hotter than the fully post-superoutburst
disk in WZ Sge-type dwarf novae and there may be remaining
pressure effect.  Since stage C superhumps often continue 
to exist after the termination of the superoutburst, 
we nevertheless estimated the disk radius 
by the same method used above. 
The results are shown in table \ref{tab:rdiskfromstagec}
and figure \ref{fig:qrstagec}.  We note here that these results 
should be looked at with a caveat mentioned above. 
The majority of the objects
have a disk radius 0.35$\pm$0.04$A$, which is not very
different from that estimated from post-superoutburst superhumps
in WZ Sge-type dwarf novae.  By using this assumption, we may
estimate $q$ for objects without measurements of stage A
superhumps with an uncertainty of about $\pm$20\%.
An object with a large deviation from the general trend
(V592 Her) can be understood as a result of poor observation
during the late stage.

\begin{table}
\caption{Disk radii estimated from the periods of stage C superhumps.}
\label{tab:rdiskfromstagec}
\begin{center}
\begin{tabular}{cccc}
\hline
Object & Year & $R_{\rm disk}$ & error \\
\hline
V466 And & 2008 & 0.322 & 0.005 \\
VY Aqr & 2008 & 0.335 & 0.004 \\
WX Cet & 1989 & 0.322 & 0.019 \\
WX Cet & 1998 & 0.338 & 0.007 \\
V632 Cyg & 2008 & 0.392 & 0.004 \\
UV Per & 2003 & 0.313 & 0.002 \\
SW UMa & 2006 & 0.347 & 0.003 \\
KS UMa & 2003 & 0.382 & 0.002 \\
HV Vir & 2002 & 0.379 & 0.005 \\
1RXS J0423 & 2008 & 0.380 & 0.001 \\
ASAS J0233 & 2006 & 0.348 & 0.011 \\
ASAS J1025 & 2006 & 0.377 & 0.002 \\
ASAS J1600 & 2005 & 0.299 & 0.002 \\
SDSS J1556 & 2007 & 0.387 & 0.003 \\
V592 Her & 2010 & 0.240 & 0.038 \\
IY UMa & 2009 & 0.361 & 0.002 \\
OT J1044 & 2010 & 0.394 & 0.006 \\
V1504 Cyg & 2009b & 0.346 & 0.001 \\
V344 Lyr & 2009 & 0.381 & 0.001 \\
V344 Lyr & 2009b & 0.367 & 0.001 \\
SW UMa & 2010 & 0.404 & 0.005 \\
OT J210950 & 2011 & 0.342 & 0.003 \\
\hline
\end{tabular}
\end{center}
\end{table}

\subsection{ER UMa Stars and Implication to TTI Model}

   So far, none of ER UMa stars, a subgroup of SU UMa stars
with very frequent outbursts and short supercycles
(\cite{kat95eruma}; \cite{rob95eruma}; \cite{pat95v1159ori}),
has been reported to show stage A superhumps in particular 
for those systems with a low mass ratio $q$ such as 
ER UMa and V1159 Ori.  
This can be understood in the framework of the TTI model
as follows.  In ER UMa stars, the superoutburst is not triggered
by a normal outburst as in many SU UMa-type dwarf novae,
but the 3:1 resonance starts to appear before the onset
of the superoutburst (``Case C'' superoutburst in
\cite{osa03DNoutburst}, \cite{osa05DImodel}) because 
the disk becomes larger than the 3:1 resonance even in quiescence.  
In this case, the tidal instability triggers the superoutburst.
In such a condition, the superhump wave is not restricted
to the 3:1 resonance radius at the onset of the superoutburst,
and the pressure effect is expected to be already strong,
i.e. the object is already in stage B in the early phase
of the superoutburst.  Although true stage A may be found when
the object is still in quiescence, it may be difficult to
detect because the luminosity of the disk is much lower
than in the superoutburst.

   The result that the $q$ values from stage A superhumps
very well reproduces the $q$ values from quiescent eclipse
observations verifies that our assumption that the disk
radius is close to the 3:1 resonance when the superhumps
start to emerge.  This is the very consequence of the TTI
model, and the present result (and the lack of stage A
superhumps in ER UMa stars) again strengthens the TTI model.

\subsection{Quiescent Superhumps in AL Com and Other Systems}

   \citet{abb92alcomcperi} detected a very long-period
superhump period (89.6 min = 0.0622~d) in AL Com in quiescence.  
This period has $\varepsilon^*$=0.089, and \citet{pat96alcom}
suggested such a large $\varepsilon^*$ could arise from
the disk close to the 2:1 resonance.  We consider this case.
Since the $q$ value is not well known in AL Com,
we used $q$=0.078 of WZ Sge, which has very similar $P_{\rm orb}$
and $P_{\rm SH}$ to AL Com.  Because the pressure effect
can be neglected in the quiescent disk, we can use the method
in subsection \ref{sec:diskradiuspostSO}.  We could obtain
$\varepsilon^*$=0.089 at a disk radius of 0.66$A$.  This is
close to the radius for the 2:1 resonance (0.61$A$ for $q$=0.078).
Considering the uncertainty in $q$, this result seems to
support the interpretation by \citet{pat96alcom}.

   Similar signals, although they may have not been as obvious
to be directly seen in the raw light curve as in AL Com,
have been reported in other systems: BW Scl ($\varepsilon^*$=0.104,
\cite{uth12j1457bwscl}), possibly EQ Lyn ($\varepsilon^*$=0.13,
\cite{szk10CVWDpuls}) and V455 And [$\varepsilon^*$ may be 0.084
if one-day alias is allowed for the data in \citet{ara05v455and},
\citep{szk10CVWDpuls}].
Using our $q$=0.067 for BW Scl, the large $\varepsilon^*$ requires a disk
radius of 0.70$A$.  Even allowing $q$=0.073, it requires 0.69$A$.
The value for EQ Lyn is even larger, and these (possibly transient)
phenomena may be different from superhumps.

\section{Summary}

   We have examined an assumption that the superhump period
during the early growing stage of superhumps (stage A superhumps)
reflects the dynamical precession rate at the 3:1 resonance,
a picture  introduced in \citet{osa13v344lyrv1504cyg}.
We have found that the $q$ values estimated from this assumption
and $q$ values from quiescent eclipse observations are
in good agreement.  This led to $q$ estimation of a number of
SU UMa-type dwarf novae without measured $q$ values.
The obtained $q$ values followed the same CV evolutionary sequence
in $P_{\rm orb}-q$ diagram determined from quiescent 
eclipse observations in the literature,
and we consider this agreement strengthens the validity of our
method.  Our method gave systematically larger $q$ for
short-$P_{\rm orb}$ systems (WZ Sge-type dwarf novae)
compared to \citet{pat11CVdistance}, which we interpret as a result
of the stronger pressure effect in low-$q$ systems in his case.
This difference indicates that the most of the secondaries of
known WZ Sge-type dwarf novae are near the border between
lower main-sequence and brown dwarfs,
and the mass of the secondary at the period minimum is higher than
in \citet{pat11CVdistance}.  Using our $q$ values, we provide
new experimental formulae to convert the observed period excess 
of fully grown superhumps (stage B superhumps) to $q$.
This method was also used to estimate the disk radius after
the termination of superoutburst.

   Our study particularly suggests that estimation of $\varepsilon^*$ 
from observations of stage A superhumps can rival the accuracy
of the quiescent eclipse observations, and the estimated $q$
from stage A superhumps may be even considered as accurate
as $q$ estimates by other dynamical methods.  
This leads to a very important
suggestion to future observations: since stage A lasts 
only for 1--2~d, it is crucial to start observing
superhumps immediately following the onset of the superoutburst,
and multi-longitudinal observations are indispensable to record
a phase lasting only for 1--2~d.  Such a requirement for
observation naturally explains why ``stage A superhump'' only
became apparent in this modern era: it is a direct outcome
of world-wide cooperation of amateur-professional observers 
and immediate announcements of outburst detections via internet
(no outbursts have been kept secret!), 
both of which we, especially the VSNET Collaboration \citep{VSNET},
have been striving for.  Our present research predicts that
more coordinated immediate notifications and intensive
multi-longitudinal observations in the very early stage of
already known and newly discovered SU UMa-type dwarf novae
will bring many more $q$ estimations to life
whose number can easily surpass the past $q$ estimates of DNe
obtained by analyzing quiescent eclipses or radial-velocity
studies mobilizing huge telescopes.  It will certainly enrich
our understanding of the still poorly known late stage of
CV evolution.  And this is now feasible with distributed small
telescope around the globe -- why don't you proceed in this way!

\medskip

   We express our thanks to a discussion with Prof. S. Kato
for the relative strength of the pressure and dynamical
effects.

\end{document}